# Co-Creation of Innovative Gamification Based Learning: A Case of Synchronous Partnership

Nicholas Dacre [a], Vasilis Gkogkidis [a*], & Peter Jenkins [b]

[a] Science Policy research Unit, University of Sussex Business School, University of Sussex, Brighton, UK
[b] Brighton Business School, University of Brighton, Brighton, UK
[*] Corresponding Author: v.gkogkidis@sussex.ac.uk

## Abstract

In higher education, gamification offers the prospect of providing a pivotal shift from traditional asynchronous forms of engagement, to developing methods to foster greater levels of synchronous interactivity and partnership between and amongst teaching and learning stakeholders. The small vein of research that focuses on gamification in teaching and learning contexts, has mainly focused on the implementation of pre- determined game elements. This approach reflects a largely asynchronous approach to the development of learning practices in educational settings, thereby limiting stakeholder engagement in their design and adoption. Therefore, we draw on the theory of co-creation to examine the development process of gamification-based learning as a synchronous partnership between and amongst teaching and learning stakeholders. Empirical insights suggest that students gain a greater sense of partnership and inclusivity as part of a synchronous co-creation gamification-based learning development and implementation process.

**Keywords**: Co-Creation Theory, Gamification, Engagement, Partnership, Higher Education, Synchronous, Innovative, Students, Learning, Pedagogy, University, Asynchronous.

## Introduction

Coates (2005) argues that student engagement is an important indicator of the quality of a university course and it should be measured and factored in when trying to improve teaching methods and processes. The role of partnerships in fostering engagement has been linked to pivotal higher education factors such as the student experience, and teaching and learning practice (Bryson, 2016; Bryson & Hand, 2007). Research suggests an array of tools, methods, and concepts upon which pedagogic practice can draw on to elicit novel, and innovative approaches to engage students (Kirkwood & Price, 2014; McLaughlin et al., 2014). One such approach which has steadily been gaining interest and relevance with learning and teaching practice, is through the concept of gamification (Dacre, Constantinides, & Nandhakumar, 2015; Tsay & Kofinas, 2017; Urh, Vukovic, & Jereb, 2015).

Gamification is defined by Deterding et al. (2011, p. 9) as "the use of game design elements in non-game contexts". Research suggests that the application of game-based elements has the potential to enhance engagement and promote participative interaction (Dacre, Constantinides, & Nandhakumar, 2015; Nacke & Deterding, 2017).

As suggested by Fotaris et al. (2016), gamification can be an effective tool when successfully implemented to help increase student engagement with teaching and learning practice. However, the majority of gamification research in higher education has mainly focused on introducing pre-designed gamified solutions to understand their influence on practice (Iosup & Epema, 2014; Tsay & Kofinas, 2017; Urh et al., 2015). In these settings the level of synchronous partnership is limited by the asynchronous gamification development, whereby educators implement pre-designed solutions. Conversely, the concept of co-creation offers a much more dynamic and interactive partnership during the process of product and service development.

The idea of introducing co-creation in the context of higher education was explored by Díaz-Méndez and Gummesson (2012) suggesting a positive influence on student engagement. Co-creation is described by Vargo and Lusch (2008) and Barile and Polese (2010) as the joint creation of value where customers and organisation work together both offering resources to improve the outcome of the design process. Therefore, this paper draws on the theory of co-creation to examine the partnership in developing a gamification-based learning system to improve engagement amongst and between teaching and learning stakeholders on a postgraduate course.

The programme that the gamification-based learning system is designed for is an elective postgraduate business management module. It is an intensive programme that teaches students when and how to use gamification elements in business contexts. Our findings suggest that students gain a greater sense of partnership and inclusivity as part of a synchronous co-creation development and implementation process as well as high levels of satisfaction from the course.

## Research Approach

The gamification-based learning system that was co-created by the teaching and learning stakeholders, included game design elements like missions, virtual currency, power ups and virtual treasure chests (Dacre, Constantinides, & Nandhakumar, 2015; Huang & Soman, 2013).

The co-creation process was undertaken with all stakeholders by clearly agreeing to the goals and expected outcomes of the system, responding to suggestions through regular feedback sessions, and experimenting with new gamification elements. At the end of each day a short feedback session would be facilitated by the lecturers reflecting on the gamification system and how the learnings of that specific day might influence it. Students needed to complete missions to win virtual coins that would allow them to buy power ups and treasure chests. Treasure chests offered various rewards to the students. Missions that were co-created as part of the gamification development process, where players could earn coins upon completion included:

| |
|---|
| Not exceeding the time given to present |
| Not exceeding the number of slides given to present |
| Best presentation of the day (that was decided through voting that included only students) |
| Answer quiz questions during an everyday Kahoot session. |
| Present without slides |
| Present without talking |

The power ups that were co-created where students could buy with the coins they got from completing missions were:

| |
|---|
| Give someone two more minutes for their next presentation |
| Take away two minutes from someone's presentation |
| Increase the number of slides on someone's presentation by two |
| Decrease the number of slides on someone's presentation by two |
| Someone from the class has to wear a funny hat the whole day |

A Google spreadsheet was created to facilitate the co-creation process, where all students had access to view a dashboard and review suggestions, developments, how many earned coins, what they can buy with their coins, and how many times they completed each mission. The spreadsheet was updated on a daily basis with the latest developments and results from the day's activities and record of students' progress and engagement.

Data collection also comprised of a combination of semi-structured interviews and focus groups where students offered feedback and further design ideas (Denzin & Lincoln, 1994; Neuman, 2013).

| | Silver Coins | Gold Coins |
|---|---|---|
| | 250 | 813 |
| **Missions** | | |
| Keep the time | 250 | |
| Use the right number of slides | 250 | |
| Have the best presentation | 250 | |
| Kahoot | Kahoot points/10 | 450 |

Figure 1: Example of the Google spreadsheet that kept track of the student engagement.

# Findings

Co-creation resulted in a very fruitful and productive partnership between the students and the lecturers (Bryson, 2016; Díaz-Méndez & Gummesson, 2012). Students were able to achieve 100% success rate in passing the module and when asked if being part of the design process helped them pass the module, 100% answered:

> "Yes, because I understood more about how gamification works"

Additional data from the university's standard survey asking for feedback about the course further supported the belief that including students in the design process improves engagement and satisfaction. When asked if they would recommend this course to their friends 100% answered "Definitely would" and some students even said:

> "It has been definitely my favourite module out of the ones we have been studying this year"

There were a number of salient elements of the gamification-based learning system which were interpreted as being more successful (Dacre, Constantinides, & Nandhakumar, 2015). The 'give someone two more minutes' and 'give someone two more slides' power ups made complete sense to all students as they had the chance to better prepare for their final presentation. One comment that stood out was:

> "I felt that I could break the rules of the game and get more out of it"

The 'funny hat' power up that gave students the power to make someone wear a funny hat for the whole day was also enjoyed by most of the students as it 'created an atmosphere of fun in the classroom and helped break the barriers between lecturers and students'. Conversely there were elements of the gamification system which were interpreted as being less relevant through the co-creation process.

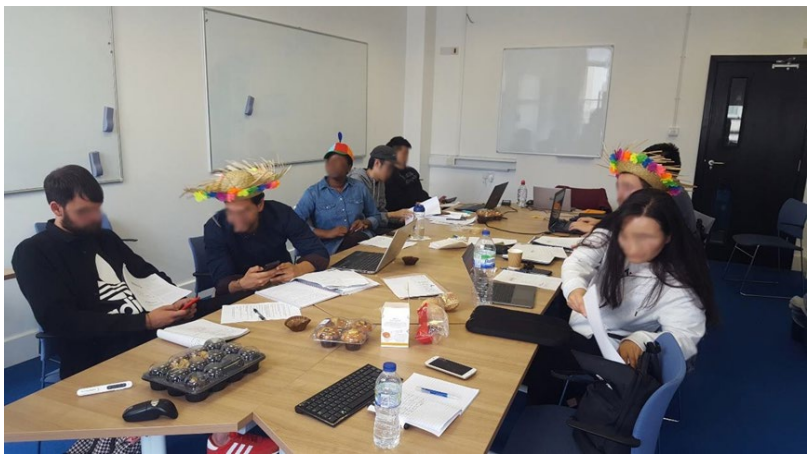

Figure 2: Students enjoying the "funny hat" power up and coming up with ideas about the gamification learning system.

Feedback was unanimous on the "present without slides" and "present without talking" missions. They initially were incorporated into the design of the system as they were deemed fun and challenging but were not aligned with the goals students had for the course. The following comment reflected the overall consensus:

> "The final presentation includes both talking and using slides, so why would I go for any of these missions when they actually take away from final goal that is to deliver a good presentation at the end of the course?"

## Conclusion

The use of a co-creation approach to embedding a gamification-based learning system, resulted in higher level of engagement and a sense of partnership. In the student's own words:

> "I felt like we were working on something that will help future students and at the same time we were learning how to improve it"

Future work should be focused on building and testing new versions of gamification learning systems using the co- creation methodology while getting data on student excellence and levels of satisfaction from the course.